\documentclass[epj]{svjour}
\usepackage[latin1]{inputenc}
\usepackage{alltt,amsmath,amssymb,colordvi,graphicx,pifont}

\newcommand{\eg}{e.g.\ }
\newcommand{\ie}{i.e.\ }
\newcommand{\bM}{\mathbf{M}}
\newcommand{\cM}{\mathcal{M}}
\newcommand{\ri}{\mathrm{i}}

\newcommand{\re}{\mathrm{e}}
\newcommand{\limfunc}[1]{\mathop{\mathrm{#1}}}
\renewcommand{\Re}{\limfunc{Re}}
\renewcommand{\Im}{\limfunc{Im}}
\newcommand{\fh}{FeynHiggs~}
\newcommand{\GeV}{~\mathrm{GeV}}

\newcommand{\Code}[1]{\ensuremath{\text{\tt #1}}}
\newcommand{\Var}[1]{\ensuremath{{\it #1}}}
\newcommand{\lbrac}{\symbol{123}}
\newcommand{\rbrac}{\symbol{125}}
\newcommand{\Brac}[1]{\lbrac#1\rbrac}
\newcommand{\eff}{\mathrm{eff}}
\newcommand{\aeff}{\ensuremath{\alpha_\eff}}

\newcommand{\BR}{\limfunc{BR}}
\newcommand{\Sqrt}[1]{\sqrt{\vphantom{g}#1}}

\newcommand\asat{\text{\ding{192}}}
\newcommand\twoloop{\text{\ding{193}}}
\newcommand\oneloop{\text{\ding{194}}}
\newcommand\order[1]{\ensuremath{\mathcal{O}(#1)}}

\usepackage{graphicx}
\usepackage{fancyhdr}

\def\mytitle{My title} 
\def\myauthors{My name}  
\def\mytype{My type of session}
\def\mysession{My session}

\begin{document}

\graphicspath{{figs/}}

\def\mytitle{Higgs Masses and More in the cMSSM with \fh}
\title{Higgs Masses and More in the Complex MSSM \\ with \fh}

\def\mytype{Parallel Session}
\def\mysession{Higgs}

\def\myauthors{Hahn, Heinemeyer, Hollik, Rzehak, Weiglein}
\author{%
  T.~Hahn\inst{1}\thanks{\emph{e-mail:} hahn@feynarts.de},
  S.~Heinemeyer\inst{2},
  W.~Hollik\inst{1},
  H.~Rzehak\inst{3}\thanks{\emph{e-mail:} Heidi.Rzehak@psi.ch},
  G.~Weiglein\inst{4}}


\institute{%
  Max-Planck-Institut f\"ur Physik,
  F\"ohringer Ring 6,
  D--80805 M\"unchen, Germany
\and
  Instituto de Fisica de Cantabria (CSIC-UC),
  Santander, Spain
\and
  Paul Scherrer Institut,
  W\"urenlingen und Villigen,
  CH--5232 Villigen PSI, Switzerland
\and
  IPPP, University of Durham,
  Durham DH1~3LE, UK}

\date{}

\abstract{%
We present the latest version 2.6 of FeynHiggs, a program for computing MSSM
Higgs-boson masses and related observables, such as mixing angles, branching 
ratios, and couplings, including state-of-the-art higher-order contributions.  
The most important new feature is the inclusion of the fully complex 
\order{\alpha_t\alpha_s} two-loop corrections, which enables \fh to give
the most 
precise Higgs-mass evaluation in the complex MSSM in the
Feynman-diagrammatic approach to date.
}

\maketitle


\section{Complex Parameters in the MSSM Higgs Sector}

The Higgs sector of the Minimal Supersymmetric Standard Model with
complex parameters (cMSSM) consists of two Higgs doublets
\begin{align}
H_1 &= \begin{pmatrix}
       v_1 + \frac 1{\sqrt 2}(\phi_1 - \ri\chi_1) \\
       -\phi_1^-
       \end{pmatrix}, \\
H_2 &= \Magenta{\re^{\ri\xi}}
       \begin{pmatrix}
       \phi_2^+ \\
       v_2 + \frac 1{\sqrt 2}(\phi_2 + \ri\chi_2)
       \end{pmatrix}
\end{align}
which form the following Higgs potential
\begin{align}
V &= m_1^2\,H_1\bar H_1 + m_2^2\,H_2\bar H_2 -
     (\Magenta{m_{12}^2}\,\varepsilon_{\alpha\dot\beta}
       H_1^\alpha H_2^{\dot\beta} + \mathrm{h.c.}) \notag \\
&\quad + \frac{g_1^2 + g_2^2}{8}\,
       (H_1\bar H_1 - H_2\bar H_2)^2 +
     \frac{g_2^2}{2}\,|H_1\bar H_2|^2.
\end{align}
The Higgs potential contains two complex phases \Magenta{$\xi$},
\Magenta{$\arg(m_{12}^2)$}.  
The phase \Magenta{$\arg(m_{12}^2)$} can be rotated away
\cite{Peccei,MSSMcomplphasen} and, at 
tree level, \Magenta{$\xi$} has to vanish in order to fulfill the
minimum condition of the Higgs potential, so there
is no CP-violation at tree level and the
spectrum contains five states of definite CP-parity: $h$, $H$, $A$, $H^\pm$.
In the following we review the inclusion of higher-order corrections to
Higgs-boson masses and more into the code
FeynHiggs~\cite{feynhiggs,mhiggslong,mhiggsAEC,mhcMSSMlong}. 

CP-violating effects are induced by complex parameters that enter via
loop corrections: the Higgsino mass parameter $\mu$, the trilinear
couplings $A_{t,b,\tau}$, and the gaugino mass parameters $M_{1,2,3}$. 
They yield $\Magenta{\hat\Sigma_{hA}}$, $\Magenta{\hat\Sigma_{HA}}\neq 0$ 
and induce mixing between $h$, $H$, and $A$~\cite{mhiggsCPV}.
The Higgs mass matrix has the form
\begin{small}
\begin{equation}
\bM^2 = \begin{pmatrix}
q^2 - m_h^2 + \hat\Sigma_{hh} &
        \hspace*{1em}\hat\Sigma_{hH} &
                \hspace*{1em}\Magenta{\hat\Sigma_{hA}} \\[.4ex]
\hat\Sigma_{Hh}\hspace*{1em} &
        \hspace*{-1.9em}
        q^2 - m_H^2 + \hat\Sigma_{HH}
        \hspace*{-1.9em} &
                \hspace*{1em}\Magenta{\hat\Sigma_{HA}} \\[.4ex]
\Magenta{\hat\Sigma_{Ah}}\hspace*{1em} &
        \Magenta{\hat\Sigma_{AH}}\hspace*{1em} &
                q^2 - m_A^2 + \hat\Sigma_{AA}
\end{pmatrix}\!,
\end{equation}
\end{small}%
where $m_{h,H,A}$ denote the tree-level Higgs masses, and 
it should be noted that in general $\bM^2$ is symmetric but not 
Hermitian.  In the approximation of vanishing external momentum ($q^2 = 
0$), one can obtain the higher-order corrected mass eigenstates via a 
unitary transformation from the tree-level states:
\begin{equation}
\label{def:U}
\begin{pmatrix}
  h_1 \\ h_2 \\ h_3
\end{pmatrix} = \begin{pmatrix}
  U_{11} & U_{12} & \Magenta{U_{13}} \\
  U_{21} & U_{22} & \Magenta{U_{23}} \\
  \Magenta{U_{31}} & \Magenta{U_{32}} & U_{33}
\end{pmatrix} \begin{pmatrix}
  h \\ H \\ A
\end{pmatrix}.
\end{equation}


\section{Higgs-boson self-energy corrections in \fh}

\subsection{Higgs-boson masses}

The following contributions to the mass matrix and the 
charged-Higgs-boson self-energy are taken into account:
\begin{gather}
\begin{pmatrix}
q^2 - m_h^2 + \hat\Sigma_{hh}^{\asat\twoloop\oneloop} &
        \hspace*{1.1em}\hat\Sigma_{hH}^{\asat\twoloop\oneloop} &
             \hspace*{1.2em}\Magenta{\hat\Sigma_{hA}^{\asat\oneloop}} \\[.4ex]
\hat\Sigma_{Hh}^{\asat\twoloop\oneloop}\hspace*{1.2em} &
        \hspace*{-1.9em}
        q^2 - m_H^2 + \hat\Sigma_{HH}^{\asat\twoloop\oneloop}
        \hspace*{-1.9em} &
             \hspace*{1.2em}\Magenta{\hat\Sigma_{HA}^{\asat\oneloop}} \\[.4ex]
\Magenta{\hat\Sigma_{Ah}^{\asat\oneloop}}\hspace*{1.2em} &
        \Magenta{\hat\Sigma_{AH}^{\asat\oneloop}}\hspace*{1.1em} &
                q^2 - m_A^2 + \hat\Sigma_{AA}^{\asat\oneloop}
\end{pmatrix}, \notag \\
\hat\Sigma_{H^+H^-}^{\asat\oneloop}
\end{gather}

\begin{itemize}
\item[\asat]
Leading \order{\alpha_t\alpha_s} cMSSM two-loop corrections 
\cite{asatcplx}.

\item[\twoloop]
Leading \order{\alpha_t^2} and subleading \order{\alpha_b\alpha_s,
\alpha_t\alpha_b, \alpha_b^2} two-loop corrections evaluated in the
MSSM with real parameters (rMSSM), where the phases are included only
partially \cite{atat,asab,atab}.

\item[\oneloop]
Full one-loop evaluation (all phases, $q^2$ dependence)
\cite{mhcMSSMlong} and leading non-minimal flavour-violating \linebreak
(NMFV) corrections \cite{NMFV}. 
\end{itemize}
\fh performs a numerical search for the complex roots of
$\det\bM^2(q^2)$ which are denoted as $\cM^2_{h_i}$, $i = 1\dots 3$.  A
decomposition can be performed,
\begin{align}
\cM^2 = M^2 - i M \Gamma\,,
\end{align}
where $M$ is the mass of the particle and $\Gamma$ its width.  We then 
define the loop-corrected masses according to
\begin{align}
M_{h_1} \leqslant M_{h_2} \leqslant M_{h_3}\,.
\end{align}
The Higgs masses are thus determined as the real parts of the complex
poles of the propagator.  Complex contributions to the Higgs mass matrix
(from $\Im\hat\Sigma$) are taken into
account~\cite{mhcMSSMlong,lcws07FH}.  The diagonalization routines are
available as a stand-alone package from the Web site
www.feynarts.de/diag \cite{diag}.

\subsection{Two-loop corrections in the complex MSSM}

Including the phase dependence, the complete one-loop~\cite{mhcMSSMlong}
and the two-loop contribution of
\order{\alpha_t\alpha_s}~\cite{asatcplx} to the Higgs self-energies are
taken into account.  Within the Higgs sector, the parameters have to be
defined up to \order{\alpha_t\alpha_s}.  The masses of the charged
Higgs boson, the $Z$-boson, as well as the $W$-boson are defined as pole
masses,
\begin{align}
\delta {M_X^2}^{(i)} = \text{Re} \Sigma^{(i)}_{XX} (M^2_X) \;
\text{with} \; X = 
\{H^\pm,\, W,\, Z\}, 
\end{align}
with $(i)$ denoting the loop order.  Furthermore, it is required that
there be no shift of the minimum of the Higgs potential which is fixing 
the tadpole parameters,
\begin{align}
\delta t^{(i)}_\phi = - T^{(i)}_\phi
\quad\text{with}\quad
\phi = \{h, H, A\}\,.
\end{align}
The Z-factors and $\tan\beta$ are defined within the
$\overline{\text{DR}}$-scheme~\cite{DRbarOS,doinkDRbarOS}.

The parameters of the top sector have to be defined at one-loop
level.  The top-quark mass and the top-squark masses are fixed by an
on-shell condition and the mixing angle and the corresponding phase
by
\begin{align}
\widetilde{\Re}\hat\Sigma_{\tilde t_{12}}(m^2_{\tilde t_1}) +
\widetilde{\Re}\hat\Sigma_{\tilde t_{12}}(m^2_{\tilde t_2}) = 0\,,
\end{align}
generalizing the renormalization conditions imposed in \cite{hr} for the
use of complex parameters.

To extract the relevant terms at two-loop order we used the
approximation of vanishing external momenta and vanishing electroweak
gauge couplings in the evaluation of all two-loop diagrams including
those needed for calculating the two-loop counterterms. 

\medskip

A new flag in \fh (see Sec.~\ref{sec:commandline} below) controls
the treatment of phases in the part of the two-loop corrections known
only in the rMSSM so far.  The following options are possible:
\begin{itemize}
\item all corrections: \order{\alpha_t\alpha_s, \alpha_b\alpha_s,
  \alpha_t^2, \alpha_t\alpha_b, \alpha_b^2} in the rMSSM,
\item only the cMSSM \order{\alpha_t\alpha_s} corrections,
\item the cMSSM \order{\alpha_t\alpha_s} corrections combined with
  the remaining corrections in the rMSSM, truncated in the phases,
\item the cMSSM \order{\alpha_t\alpha_s} corrections combined with
  the remaining corrections in the rMSSM, interpolated in the phases
  [default].
\end{itemize}
\fh thus not only has the most precise evaluation of the Higgs masses in
the cMSSM available to date (using the Feynman-diagrammatic approach),
but also a method to obtain a reasonably objective estimate of the
uncertainties due to the rMSSM-only parts.

Implementing the \order{\alpha_t\alpha_s} cMSSM corrections in \fh was a
major piece of work.  The amplitudes could be shrunk from 38 MB to less
than 1.5 MB, mainly by abbreviationing techniques and exploiting the
unitarity of the sfermion mixing matrices.  The compile time is about 3
min (up from 45 sec in \fh 2.5) and the run time is 28 msec per
parameter point (up from 27 msec in \fh 2.5).  These figures show that
the full cMSSM evaluation is actually usable in everyday life.

\medskip

\begin{figure}[b!]
\begin{center}
\includegraphics[width=\hsize]{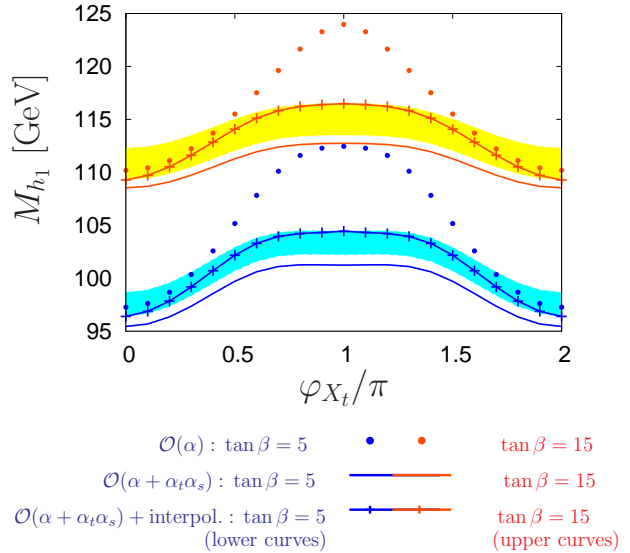}
\end{center}
\vspace{-1em}
\caption{The lightest cMSSM Higgs-boson mass as a function of
  $\varphi_{X_t}$ at the one- and two-loop level (see text).  The other
  parameters are:
$M_{\mathrm{SUSY}} = M_3 = M_2 = 500\GeV$, $M_1 = 250\GeV$,
$\mu = 1000\GeV$,
$M_{H^\pm} = 150\GeV$,
$|X_t| = 700\GeV$.}
\label{fig:Mh1phiXt}
\vspace{-1em}
\end{figure}

As a numerical example we show in Fig.~\ref{fig:Mh1phiXt} the dependence
of $M_{h_1}$ on the phase of the top-squark mixing, $X_t = A_t -
\mu^* \cot\beta$.  The plot shows the one-loop-corrected Higgs mass
$M_{h_1}$ as dotted curve.  The drawn-through curve depicts the Higgs
mass $M_{h_1}$ including contributions of \order{\alpha_t\alpha_s}. 
The boundaries of the bands are calculated in the following way:
\begin{align}
M_{h_1}^{\text{low}}(\varphi_{X_t}) =
  M_{h_1}^{\text{corr.}}(\varphi_{X_t}) +
  \Delta M_{h_1}(\varphi_{X_t} = 0)\,, \label{eq:low} \\
M_{h_1}^{\text{up}}(\varphi_{X_t}) =
  M_{h_1}^{\text{corr.}}(\varphi_{X_t}) +
  \Delta M_{h_1}(\varphi_{X_t} = \pi)\,, \label{eq:up}
\end{align}
where $M_{h_1}^{\text{low}}$ and $M_{h_1}^{\text{up}}$ respectively give 
the lower and the upper boundary of the bands for 
\begin{align}
\label{eq:deltams}
\Delta M_{h_1}(\varphi_{X_t} = 0) \leqslant
  \Delta M_{h_1}(\varphi_{X_t} = \pi)\,.
\end{align}
If, unlike in our numerical example, Eq.~\eqref{eq:deltams} does not
hold, $M_{h_1}^{\text{low}}$ and $M_{h_1}^{\text{up}}$ have to be
interchanged in Eqs.~\eqref{eq:low} and \eqref{eq:up}.

$M_{h_1}^{\text{corr.}}$ are the values for $M_{h_1}$ including the full
one-loop and the \order{\alpha_t\alpha_s} corrections with the full
phase dependence.  $\Delta M_{h_1}$ gives the size of the contributions
that are only known for real parameters, namely those of
\order{\alpha_t^2, \alpha_b\alpha_s, \alpha_t\alpha_b, \alpha_b^2}.

The crossed curve shows $M_{h_1}$, taking into account the
\order{\alpha_t\alpha_s} contributions and interpolating $\Delta
M_{h_1}$, \ie the corrections of \order{\alpha_t^2, \alpha_b\alpha_s,
\alpha_t\alpha_b, \alpha_b^2}.  The fact that these crossed curves 
lie between the lower and the upper boundaries of the corresponding band
shows that the interpolation procedure is working well.

For the parameters chosen here (especially due to a relatively small
value of $M_{H^\pm}$) the \order{\alpha_t\alpha_s} contributions
decrease the phase dependence.  As known from the
rMSSM~\cite{mhiggslong}, they cause a 
shift of $M_{h_1}$ towards lower values with respect to the one-loop
corrected mass, the \order{\alpha_t^2, \alpha_b\alpha_s,
\alpha_t\alpha_b, \alpha_b^2} corrections increase again the size of the
$M_{h_1}$.

\subsection{Mixing of the Higgs bosons}

\fh returns two different `mixing' matrices.
\begin{itemize}
\item
\Code{UHiggs} is a `true' mixing matrix in the sense of being unitary 
and hence preserving probabilities. When applying effective couplings for 
internal Higgs bosons, this matrix must be used.

It should be noted that to obtain a unitary matrix, it is mathematically
a necessity that $\bM^2$ has no imaginary parts -- making it Hermitian. 
This of course constrains the achievable quality of approximation.

\item
\Code{ZHiggs} is a matrix of Z-factors.  It guarantees on-shell 
properties for external Higgs bosons~\cite{mhcMSSMlong}, see 
Eq.~(\ref{eq:Z}) below.
\end{itemize}
It is important to understand that \Code{ZHiggs} and \Code{UHiggs} are
two objects with physically and mathematically distinct properties. 
Neither is universally `better' than the other.

\Code{UHiggs} can be computed in two approximations:
\begin{itemize}
\item
$q^2$ on-shell:
$\hat\Sigma_{ii}\bigl(q^2 = m_i^2\bigr)$,
$\hat\Sigma_{ij}\bigl(q^2 = \tfrac 12 (m_i^2 + m_j^2)\bigr)$.

\item
$q^2 = 0$ (see Eq.~(\ref{def:U})).  
\Code{UHiggs} coincides with \Code{ZHiggs} in this limit and
corresponds to the effective potential approach.  In the absence of
CP-violating effects, \ie $2\times 2$ mixing only, this is identical to
the \aeff\ description~\cite{hff}.
\end{itemize}

\Code{ZHiggs} is engineered to deliver the correct on-shell properties
of an external Higgs boson, but is not necessarily unitary
\cite{mhcMSSMlong}.  The following picture shows the type of mixing
contributions which appear in the decay of an external Higgs boson
(the contributions from mixing with the Goldstone boson and with the
longitudinal component of the $Z$-boson are numerically small and hence
neglected): 
\begin{center}
\includegraphics{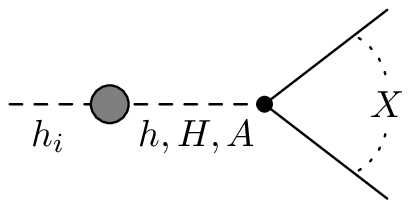}
\end{center}
Disregarding possible permutations for the moment (but see below),
the corresponding combination of amplitudes are
\begin{align}
\Gamma_{h_1} &= \Sqrt{Z_h} \bigl(
         \Gamma_h +
  Z_{hH} \Gamma_H +
  Z_{hA} \Gamma_A \bigr) \\
\Gamma_{h_2} &= \Sqrt{Z_H} \bigl(
  Z_{Hh} \Gamma_h +
         \Gamma_H +
  Z_{HA} \Gamma_A \bigr) \\
\Gamma_{h_3} &= \Sqrt{Z_A} \bigl(
  Z_{Ah} \Gamma_h +
  Z_{AH} \Gamma_H +
         \Gamma_A \bigr)
\end{align}
where
\begin{itemize}
\item $\Gamma_{h,H,A}$ is the amplitude for $h,H,A\to X$,
\item $\Sqrt{Z_{h,H,A}}$ sets the residuum of the external Higgs bosons to 1,
\item $Z_{hH}$, $Z_{hA}$ describe the transition $h\to H,A$, etc.
\end{itemize}
For convenience, the $Z$ factors can be arranged in matrix form:
\begin{equation}
\label{eq:Z}
\Code{ZHiggs} = \begin{pmatrix}
  \Sqrt{Z_h} & \Sqrt{Z_h}\,Z_{hH} & \Sqrt{Z_h}\,Z_{hA} \\[.2ex]
  \Sqrt{Z_H}\,Z_{Hh} & \Sqrt{Z_H} & \Sqrt{Z_H}\,Z_{HA} \\[.2ex]
  \Sqrt{Z_A}\,Z_{Ah} & \Sqrt{Z_A}\,Z_{AH} & \Sqrt{Z_A}
\end{pmatrix} .
\end{equation}
In this guise, \Code{ZHiggs} can be used very much like \Code{UHiggs}
even though its theoretical origin is quite different.  Reassuringly, 
\Code{ZHiggs} and \Code{UHiggs} coincide in the limit $q^2 = 0$.

The transition factors $Z_{ij}$ involve both the tree-level mass 
$m_i$ and the loop-corrected mass $\cM_i$ of each Higgs boson:
\begin{align}
Z_{ij} &= \frac
  {\hat\Sigma_{ik}(\cM_i^2)\,\hat\Sigma_{jk}(\cM_i^2) -
    \hat\Sigma_{ij}(\cM_i^2) Y_{ij}}
  {Y_{ij} Y_{ik} - \hat\Sigma_{jk}^2(\cM_i^2)}, \\
Y_{ij} &= \cM_i^2 - m_j^2 + \hat\Sigma_j(\cM_i^2)\,.
\end{align}
To compute $Z_{ij}$ we thus have to make the connection between the 
`loop' ($h_1$, $h_2$, $h_3$) and the `tree' ($h$, $H$, $A$) states.
Neither the zero-search nor the diagonalization procedure allow to do
this in an unambiguous way.  For example, level crossings may occur when
searching for the zeros of $\det\bM^2$.

The algorithm currently used by \fh is: compute \Code{ZHiggs} and the 
associated masses $\tilde M_i$ for all permutations $\pi$ of Higgs 
states involved in the mixing and choose the one which minimizes
\begin{equation}
\sum_i |M_i - \tilde M_{\pi(i)}| +
\sum_{i,j} |C_{ij} - \Code{ZHiggs}_{\pi(i)j}|
\end{equation}
where $C$ is the mixing matrix that comes out of the diagonalization of
$\bM^2$ with $q^2 = M_{h_2}^2$, \ie a by-product of the zero-search.

\begin{figure}[t!]
\begin{center}
\includegraphics[width=\hsize]{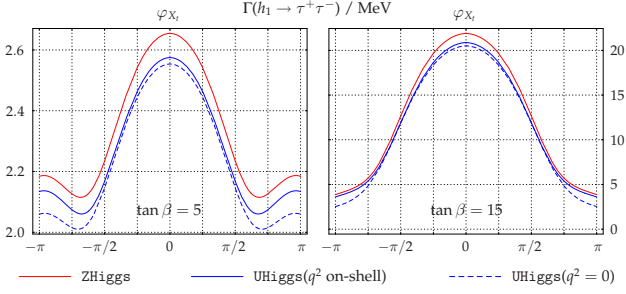}
\end{center}
\vspace{-1em}
\caption{The decay width of the lightest Higgs boson to $\tau$ leptons
  as a function of $\varphi_{X_t}$ for $\tan\beta = 5$ (left) and 
  $\tan\beta = 15$ (right).  The other parameters are:
$M_{\text{SUSY}} = M_3 = M_2 = 500\GeV$,
$\mu = 1000\GeV$,
$M_{H^\pm} = 150\GeV$,
$X_t = 700\,\re^{\ri\varphi_{X_t}}\GeV$.}
\label{fig:GaphiXt}
\vspace{-1em}
\end{figure}

This is an empirical recipe, so the different dimensions of $M$ and $Z$
should not be taken very seriously.  The permutation is decided in
nearly all cases by the mass pattern.  The $|C - Z|$ term becomes
relevant only for (almost) degenerate masses where it can tell \eg the
symmetric from the antisymmetric state.

\medskip

The numerical effects of using the various mixing matrices in a physical
amplitude are shown in Fig.~\ref{fig:GaphiXt}.  For the chosen
parameters, $\Code{UHiggs}(q^2\text{ on-shell})$ gives results closer to
the full result than $\Code{UHiggs}(q^2 = 0)$, with deviations at the
few-percent level.  For a detailed discussion see
Ref.~\cite{mhcMSSMlong}. 

\medskip

The mixing of the neutral Higgs bosons in the \linebreak cMSSM has also
been added to FeynArts \cite{feynarts}.  A special version of the MSSM
Model File \cite{famssm}, HMix.mod, provides two sets of appropriately
mixed Higgs bosons:
\begin{itemize}
\item $\Code{S[0,\,\Brac{$h$}]} =
  \sum_{i = 1}^3 \Code{UHiggs[$h$,\,$i$]}~\Code{S[$i$]}$, and
\item $\Code{S[10,\,\Brac{$h$}]} =
  \sum_{i = 1}^3 \Code{ZHiggs[$h$,\,$i$]}~\Code{S[$i$]}$.
\end{itemize}
The latter is inserted only on external lines.


\section{Benchmark Scenarios}
\label{scenarios}

\fh has long included Benchmark Scenarios \cite{LHBMS} which are useful
in the search for the MSSM Higgs bosons.  The idea is to vary only $M_A$
and $\tan\beta$ and keep all other SUSY parameters fixed.

Constraints such as Cold Dark Matter (CDM) have been ignored in these
scenarios.  It might be desirable to investigate $M_A$--$\tan\beta$
planes in agreement with CDM and other external constraints, however (if
the planes are derived in a GUT-based model, see Ref.~\cite{ehhow} for a
discussion).

For the Constrained MSSM (or mSUGRA) as candidate model, the CDM
constraints turn 
out to be too severe, \ie cut out almost all available parameter space.
This is different in the NUHM (Non-universal Higgs-mass model)
\cite{nuhm}, where the assumption is that there is no unification of
scalar fermion and scalar Higgs parameters at the GUT scale. As additional
free parameters in this model one can choose $M_A$ and $\mu$.

In Ref.~\cite{ehhow} four $M_A$--$\tan\beta$ benchmark planes have been
defined that are in agreement with the CDM and other low-energy
constraints (see also Ref.~\cite{ehoww}).  From a technical point of
view, the NUHM introduces non-trivial relations between parameters,
which thus cannot be scanned naively by independent loops.  \fh 2.6
offers the new format of Parameter Tables to deal with such cases. 

Input parameters can either be given in an input file (as in previous
versions) or interpolated from a table, in almost any mixture.  The
table format is fairly straightforward:
\begin{center}
\begin{small}
\begin{verbatim}
   MT     MSusy  MA0   TB   At     MUE ...
   170.9  500    200   5    1000   761
   170.9  500    210   5    1000   753
   ...
   170.9  500    200   6    1000   742
   170.9  500    210   6    1000   735
\end{verbatim}
\end{small}
\end{center}
For two given inputs (typically $M_A$ and $\tan\beta$) the four
neighbouring grid points are searched in the table and the other
parameters are interpolated from those points.  An error is returned if 
the inputs fall outside of the table boundaries (\ie no extrapolation).

The four predefined NUHM $M_A$--$\tan\beta$ planes \cite{ehhow} can be
obtained from www.feynhiggs.de/planes.  The definition of new planes
by the user is possible.

The Table concept is actually embedded into the new \fh Record.  This is
a data type which captures the entire content of a \fh parameter file.
Using a Record, the programmer can process \fh parameter files
independently of the frontend.


\section{Output of \fh 2.6}

We give a short overview of the output routines of the \fh library.

\bigskip

\noindent\textbf{\Code{FHHiggsCorr}}
-- All Higgs-boson masses and mixings:
$M_{h_{1,2,3}}$, $M_{H^\pm}$, $\aeff$, \Code{UHiggs}, \Code{ZHiggs}.

\medskip

\noindent\textbf{\Code{FHUncertainties}}
-- Uncertainties of the masses and mixings.

\medskip

\noindent\textbf{\Code{FHCouplings}}
-- Couplings and Branching Ratios for the following Higgs decay 
channels:
\begin{align*}
h_{1,2,3} \to {}
& f\bar f, \gamma\gamma, ZZ^{(*)}, WW^{(*)}, gg, \; &
        H^\pm \to {}
        & f^{(*)}\bar f', \\
& h_i Z^*, h_i h_j, H^+ H^-, &
        & h_i W^{\pm *}, \\
& \tilde f_i \tilde f_j, &
        & \tilde f_i \tilde f'_j, \\
& \tilde\chi_i^\pm \tilde\chi_j^\pm, \tilde\chi_i^0 \tilde\chi_j^0, &
        & \tilde\chi_i^0 \tilde\chi_j^\pm,
\end{align*}
plus the corrsponding channels of an SM Higgs with mass $M_{h_i}$:
$h_{1,2,3}^{\text{SM}}\to f\bar f, \gamma\gamma, ZZ^{(*)}, WW^{(*)}, gg.$

\medskip

\noindent\textbf{\Code{FHHiggsProd}}
-- Higgs production-channel cross-sec\-tions
(SM total cross-sections multiplied with MSSM effective couplings, see
Ref.~\cite{HiggsXS})
\begin{itemize}
\item $gg\to h_i$
      -- gluon fusion.

\item $WW\to h_i$, $ZZ\to h_i$
      -- gauge-boson fusion.

\item $W\to W h_i$, $Z\to Z h_i$
      -- Higgs-strahlung.

\item $b\bar b\to b\bar b h_i$
      -- bottom Yukawa process.

\item 
$b\bar b\to b\bar b h_i,$ -- bottom Yukawa process, one $b$ tagged.

\item $t\bar t\to t\bar t h_i$
      -- top Yukawa process.

\item $\tilde t \bar{\tilde t}\to \tilde t\bar{\tilde t} h_i$
      -- stop Yukawa process.
\end{itemize}

\noindent\textbf{\Code{FHConstraints}}
-- Electroweak precision observables, see \eg Ref.~\cite{PomssmRep} for
details:
\begin{itemize}
\item $\Delta\rho$
at \order{\alpha, \alpha\alpha_s}, including NMFV effects.

\item $M_W$, $\sin^2\theta_\eff$
via SM formula + $\Delta\rho$.

\item $\BR(b\to s\gamma)$
including NMFV effects
\cite{bsgNMFV}.

\item $(g_\mu - 2)_{\mathrm{SUSY}}$
including full one- and leading/sub\-leading two-loop SUSY corrections.

\item EDMs of electron (Th), neutron, Hg.
\end{itemize}


\section{Download and Build}

\begin{itemize}
\item
Get the \fh tar file from www.feynhiggs.de.

\item
Unpack and configure:
\begin{verbatim}
  tar xfz FeynHiggs-2.6.1.tar.gz
  cd FeynHiggs-2.6.1
  ./configure
\end{verbatim}

\item
``\Code{make}'' builds the Fortran/C++ part only.

``\Code{make all}'' builds also the Mathematica part.

The build takes about 3 min on a Pentium IV.

\item
``\Code{make install}'' installs the package.

\item
``\Code{make clean}'' removes unnecessary files.
\end{itemize}
The build was tested on Linux, Tru64 Unix, Mac OS, Windows (Cygwin) and 
also with Mathematica 6 (non-trivial due to its many incompatibilities)
and older versions.


\section{Usage}

\fh has four modes of operation:
\begin{itemize}
\item Library Mode:
Invoke the \fh routines from a Fortran or C/C++ program linked with 
\Code{libFH.a}\,.

\item Command-line Mode:
Process parameter files in \fh or SLHA format from the shell prompt or
in scripts with the \Code{FeynHiggs} stand-alone executable.

\item Web Mode:
Interactively choose the parameters at the \fh User Control Center 
(FHUCC) and obtain the results on-line.

\item Mathematica Mode:
Access the \fh routines in Mathematica via MathLink with 
\Code{MFeynHiggs}.
\end{itemize}
All programs and subroutines are documented in man pages.


\subsection{Library Mode}

The \fh library \Code{libFH.a} is a static Fortran 77 library.
Its global symbols are prefixed with a unique identifier to minimize
symbol collisions.  The library contains only subroutines (no
functions), so that no include files are needed (except for the
couplings) and the invocation from C/C++ is hassle-free.  Detailed
debugging output can be turned on at run time.  All routines are
described in detail in the API guide and on man-pages.


\subsection{Command-line Mode}
\label{sec:commandline}

The user submits a parameter (text) file, such as
\begin{center}
\begin{small}
\begin{minipage}[t]{.3\hsize}
\begin{verbatim}
MT        170.9
MA0       200
TB        50
MSusy     975
Abs(M_2)  332
Abs(MUE)  980
Abs(At)   -300
Abs(Ab)   1500
Abs(M_3)  975
\end{verbatim}
\end{minipage}
\end{small}
\end{center}
to the \fh executable with a command like
\begin{alltt}
  FeynHiggs \Var{file} [\Var{flags}]
\end{alltt}
where the \Var{flags} are optional.  The output is a human-readable 
version of the results.  Details of this (rather voluminous) output are 
tagged with a \Code{\%} and can thus be masked off with
\begin{alltt}
  FeynHiggs \Var{file} [\Var{flags}] | grep -v %
\end{alltt}
The \Code{table} utility converts the output to machine-readable
format, for example
\begin{alltt}
  FeynHiggs \Var{file} [\Var{flags}] | table TB Mh0 > \Var{outfile}
\end{alltt}
The new `\Code{table}' statement in the parameter file loads the table
(see Sect.~\ref{scenarios}) and associates two interpolation variables with it.
The changes are rather minimal:
\begin{center}
\begin{small}
\begin{tabular}[t]{l|l}
Input File ``table'' & ``inline table'' \\ \hline
\begin{minipage}[t]{.44\hsize}
\begin{verbatim}
MA0       200
TB        50
table file.dat MA0 TB
\end{verbatim}
\end{minipage} & \begin{minipage}[t]{.45\hsize}
\begin{verbatim}
MA0       200
TB        50
table - MA0 TB
MA0  TB  At    MUE ...
200  5   1000  761
210  5   1000  753
...
\end{verbatim}
\end{minipage}
\end{tabular}
\end{small}
\end{center}
Loops over parameter values (parameter scans) are possible as in former 
versions:
\begin{itemize}
\item\Code{MA0 200 400 50}, linear: 200, 250, 300, 350, 400,
\item\Code{TB 5 40 *2}, logarithmic: 5, 10, 20, 40,
\item\Code{TB 5 50 /6}, number of steps: 5, 14, 23, 32, 41, 50.
\end{itemize}


\subsection{SUSY Les Houches Accord Format}

The \Code{FeynHiggs} executable can also process files in SUSY Les
Houches Accord 2 (SLHA2) format \cite{SLHA2}.  It uses the SLHA Library
\cite{SLHALib}.  Processing of SLHA2 files can also be done in Library
Mode with the subroutine \Code{FHSetSLHA}.

\fh in fact tries to read each file in SLHA format first and if that
fails, falls back to its native format.


\subsection{Web Mode}

\begin{figure*}[htb!]
\begin{center}
\includegraphics[width=0.8\hsize]{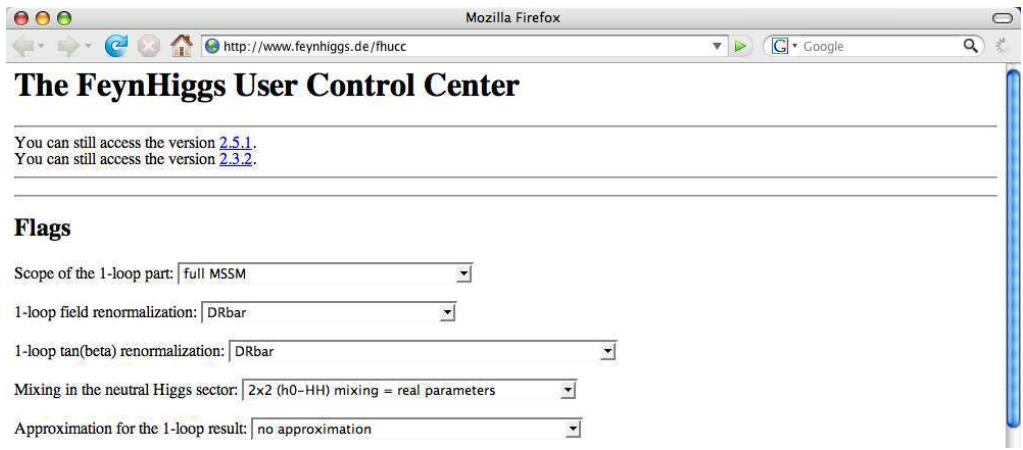}
\end{center}
\caption{Screen-shot from www.feynhiggs.de/fhucc.}
\label{fig:fhucc}
\end{figure*}

The \fh User Control Center (FHUCC) is on-line at
www.feynhiggs.de/fhucc.  It is a Web interface for the command-line
frontend.  The user gets the results together with the input file for
the command-line frontend.  A screen-shot is shown in
Fig.~\ref{fig:fhucc}.

\subsection{Mathematica Mode}

A more powerful interactive environment is provided by the Mathematica
interface of FeynHiggs.  The MathLink executable \Code{MFeynHiggs} must
first be loaded with
\begin{verbatim}
   Install["MFeynHiggs"]
\end{verbatim}
and makes all \fh routines available as Mathematica functions.  In
combination with the arsenal of standard Mathematica functions such as
\Code{ContourPlot} and \Code{Manipulate}, even sophisticated analyses
can be carried out easily.


\section{Summary: Main New Features}

Version 2.6 of \fh introduces the following new features:
\begin{itemize}
\item
Higgs masses are computed as the real part of the complex pole.

\item
Two kinds of `mixing' matrices with different properties (\Code{UHiggs}, 
\Code{ZHiggs}) are returned.  The user can choose which mixing matrix
to use in all Higgs production and decay channels (default: \Code{ZHiggs}).
   
\item
Inclusion of the full cMSSM two-loop \order{\alpha_t\alpha_s}
corrections in highly optimized form.

\item
Inclusion of full one-loop NMFV effects.

\item
Possibility to interpolate parameters from data tables.
Availability of $M_A$--$\tan\beta$ planes in agreement with 
CDM constraints.

\item 
Total Higgs production cross-sections in effective coupling
      approximation

\item
EDMs of electron (Th), neutron, Hg.
\end{itemize}



\begin{flushleft}

\end{flushleft}


\begin{thebibliography}{99}

\bibitem{Peccei}
R.~Peccei, H.~Quinn,
\textsl{Phys.\ Rev.\ Lett.} \textbf{38} (1977) 1440;
\textsl{Phys.\ Rev.} \textbf{ D16} (1977) 1791.

\bibitem{MSSMcomplphasen}
S.~Dimopoulos, S.~Thomas,
\textsl{Nucl.\ Phys.} \textbf{B465} (1996) 23
[hep-ph/9510220].

\bibitem{feynhiggs}
S.~Heinemeyer, W.~Hollik, G.~Weiglein,
\textsl{Comput.\ Phys.\ Comm.} \textbf{124} (2000) 76
[hep-ph/9812320]. 
The \fh Web site is at www.feynhiggs.de.

\bibitem{mhiggslong}
S.~Heinemeyer, W.~Hollik, G.~Weiglein,
\textsl{Eur.\ Phys.\ J.} \textbf{C9} (1999) 343
[hep-ph/9812472].

\bibitem{mhiggsAEC}
G.~Degrassi, S.~Heinemeyer, W.~Hollik, P.~Slavich, G.~Weiglein,
\textsl{Eur.\ Phys.\ J.} \textbf{C28} (2003) 133
[hep-ph/0212020].

\bibitem{mhcMSSMlong}
M.~Frank, T.~Hahn, S.~Heinemeyer, W.~Hollik, H.~Rzehak, G.~Weiglein,
\textsl{JHEP} \textbf{0702} (2007) 047
[hep-ph/0611326].

\bibitem{mhiggsCPV}
A.~Pilaftsis,
\textsl{Phys.\ Rev.} \textbf{D58} (1998) 096010
[hep-ph/9803297]; 
\textsl{Phys.\ Lett.} \textbf{B435} (1998) 88
[hep-ph/9805373]; \\
D.~Demir,
\textsl{Phys.\ Rev.} \textbf{D60} (1999) 055006
[hep-ph/9901389]; \\ 
S.~Choi, M.~Drees, J.~Lee,
\textsl{Phys.\ Lett.} \textbf{B481} (2000) 57
[hep-ph/0002287]; \\
A.~Pilaftsis, C.~Wagner, 
\textsl{Nucl.\ Phys.} \textbf{B553} (1999) 3
[hep-ph/9902371]; \\
M.~Carena, J.~Ellis, A.~Pilaftsis, C.~Wagner,
\textsl{Nucl.\ Phys.} \textbf{B586} (2000) 92
[hep-ph/0003180]; \\
T.~Ibrahim and P.~Nath,
\textsl{Phys.\ Rev.} \textbf{D63} (2001) 035009
[hep-ph/0008237]; 
\textsl{Phys.\ Rev.} \textbf{D66} (2002) 015005
[hep-ph/0204092]; \\ 
S. Heinemeyer,
\textsl{Eur.\ Phys.\ J.} \textbf{C22} (2001) 521
[hep-ph/0108059].

\bibitem{asatcplx}
S.~Heinemeyer, W.~Hollik, H.~Rzehak, G.~Weiglein, 
\textsl{Phys.\ Lett.} \textbf{B652} (2007) 300
arXiv:0705.0746 [hep-ph].

\bibitem{atat}
A.~Brignole, G.~Degrassi, P.~Slavich, F.~Zwirner,
\textsl{Nucl.\ Phys.} \textbf{B631} (2002) 195
[hep-ph/0112177].

\bibitem{asab}
A.~Brignole, G.~Degrassi, P.~Slavich, F.~Zwirner,
\textsl{Nucl.\ Phys.} \textbf{B643} (2002) 79
[hep-ph/0206101].

\bibitem{atab}
A.~Dedes, G.~Degrassi, P.~Slavich,
\textsl{Nucl.\ Phys.} \textbf{B672} (2003) 144
[hep-ph/0305127].

\bibitem{NMFV}
S.~Heinemeyer, W.~Hollik, F.~Merz, S.~Pe\~naranda,
\textsl{Eur.\ Phys.\ J.} \textbf{C37} (2004) 481
[hep-ph/0403228].

\bibitem{lcws07FH}
T.~Hahn, S.~Heinemeyer, W.~Hollik, H.~Rzehak, G.~Weiglein,
arXiv:0709.1907 [hep-ph].

\bibitem{diag}
T.~Hahn, physics/0607103.

\bibitem{DRbarOS}
M.~Frank, S.~Heinemeyer, W.~Hollik, G.~Weiglein,
hep-ph/0202166.

\bibitem{doinkDRbarOS} 
A.~Freitas and D.~St\"ockinger,
\textsl{Phys.\ Rev.} {\bf D66} (2002) 095014
[hep-ph/0205281].

\bibitem{hr}
W. Hollik, H. Rzehak, 
\textsl{Eur.\ Phys.\ J.} \textbf{C32} (2003) 127
[hep-ph/0305328].

\bibitem{hff}
S.~Heinemeyer, W.~Hollik, G.~Weiglein, 
\textsl{Eur.\ Phys.\ J.} \textbf{C16} (2000) 139
[hep-ph/0003022].

\bibitem{feynarts}
J.~K\"ublbeck, M.~B\"ohm, A.~Denner,
\textsl{Comput.\ Phys.\ Comm.} \textbf{60} (1990) 165; \\
T.~Hahn,
\textsl{Comput.\ Phys.\ Comm.} \textbf{140} (2001) 418
[hep-ph/0012260].

\bibitem{famssm}
T.~Hahn, C.~Schappacher,
\textsl{Comput.\ Phys.\ Comm.} \textbf{143} (2002) 54
[hep-ph/0105349].

\bibitem{LHBMS}
M.~Carena, S.~Heinemeyer, C.~E.~M.~Wagner, G.~Weiglein,
\textsl{Eur.\ Phys.\ J.} \textbf{C26} (2003) 601
[hep-ph/0202167];
\textsl{Eur.\ Phys.\ J.} \textbf{C45} (2006) 797,
[hep-ph/0511023].

\bibitem{ehhow}
J.~Ellis, T.~Hahn, S.~Heinemeyer, K.~Olive, G.~Weiglein,
to appear in \textsl{JHEP}, 
arXiv:0709.0098 [hep-ph].

\bibitem{nuhm}
J.~Ellis, K.~Olive, Y.~Santoso,
\textsl{Phys.\ Lett.} \textbf{B539} (2002) 107
[hep-ph/0204192]; \\
J.~Ellis, T.~Falk, K.~Olive, Y.~Santoso,
\textsl{Nucl.\ Phys.} \textbf{B652} (2003) 259
[hep-ph/0210205]; \\
M.~Olechowski, S.~Pokorski,
\textsl{Phys.\ Lett.} \textbf{B344} (1995) 201
[hep-ph/9407404];
V.~Berezinsky, A.~Bottino, J.~Ellis, N.~Fornengo, G.~Mignola, S.~Scopel,
\textsl{Astropart.\ Phys.} \textbf{5} (1996) 1
[hep-ph/9508249]; \\
M.~Drees, M.~Nojiri, D.~Roy, Y.~Yamada,
\textsl{Phys.\ Rev.} \textbf{D56} (1997) 276
[Erratum -- ibid.\ \textbf{D64} (1997) 039901]
[hep-ph/9701219]; \\
M.~Drees, Y.~Kim, M.~Nojiri, D.~Toya, K.~Hasuko, T.~Kobayashi,
\textsl{Phys.\ Rev.} \textbf{D63} (2001) 035008
[hep-ph/0007202]; \\
P.~Nath, R.~Arnowitt,
\textsl{Phys.\ Rev.} \textbf{D56} (1997) 2820
[hep-ph/9701301]; \\
A.~Bottino, F.~Donato, N.~Fornengo, S.~Scopel,
\textsl{Phys.\ Rev.} \textbf{D63} (2001) 125003
[hep-ph/0010203]; \\
S.~Profumo,
\textsl{Phys.\ Rev.} \textbf{D68} (2003) 015006
[hep-ph/0304071]; \\
D.~Cerdeno, C.~Mu\~noz,
\textsl{JHEP} \textbf{0410} (2004) 015
[hep-ph/0405057]; \\
H.~Baer, A.~Mustafayev, S.~Profumo, A.~Belyaev, X.~Tata,
\textsl{JHEP} \textbf{0507} (2005) 065
[hep-ph/0504001].

\bibitem{ehoww}
J.~Ellis, S.~Heinemeyer, K.~Olive, A.M.~Weber, G.~Weiglein,
\textsl{JHEP} \textbf{0708} (2007) 083
arXiv:0706.0652 [hep-ph].

\bibitem{HiggsXS}
T.~Hahn, S.~Heinemeyer, F.~Maltoni, G.~Weiglein, S.~Willenbrock,
[hep-ph/0607308], 
The SM cross-sections are taken from the Web site
maltoni.home.cern.ch/maltoni/TeV4LHC, providing also a
comprehensive list of original references.

\bibitem{PomssmRep}
S.~Heinemeyer, W.~Hollik, G.~Weiglein,
\textsl{Phys.\ Rept.} \textbf{425} (2006) 265
[hep-ph/0412214].

\bibitem{bsgNMFV}
T.~Hahn, W.~Hollik, J.I.~Illana, S.~Pe\~naranda,
hep-ph/0512315.

\bibitem{SLHA2}
P.~Skands et al.,
\textsl{JHEP} \textbf{0407} (2004) 036
[hep-ph/0311123]; \\
B.~Allanach, et al.,
hep-ph/0602198.

\bibitem{SLHALib}
T.~Hahn, hep-ph/0408283;
hep-ph/0605049.



\end{thebibliography}
\end{document}